\documentstyle[pra,aps,epsfig,array,twocolumn]{revtex}

\title{On Multipartite Pure-State Entanglement}
\author{Ashish V. Thapliyal}
\address{Department of Physics, University of
California, Santa Barbara, CA 93106, USA
email:{\tt ash@physics.ucsb.edu} }

\date{\today}

\begin{document}

\maketitle

\begin{abstract}
We show that pure states of multipartite quantum 
systems are {\em multiseparable\/} (i.e. give separable density matrices
on tracing any party) if and only if  they have a 
generalized Schmidt decomposition. Implications of this result for the 
quantification of multipartite pure-state entanglement are discussed. 
Further, as
an application of the techniques used here, we show that any purification
of a bipartite PPT bound entangled state  is {\em tri-inseparable\/}, i.e.
has none of its three bipartite partial traces separable. 
\end{abstract}


\newcommand{\upp}[1]{^{\rm \scriptscriptstyle #1}}
\newcommand{\dnn}[1]{_{\rm \scriptscriptstyle #1}}
\newcommand{\ket}[1]{| \, #1 \rangle}
\newcommand{\bra}[1]{ \langle #1 \,  |}
\newcommand{\proj}[1]{\ket{#1}\bra{#1}} 
\newcommand{\pket}[2]{ | \, #1 \rangle \otimes | \, #2  \rangle  } 
\newcommand{\pbra}[2]{ \langle #1\,  | \otimes \langle #2 \,  |}
\newcommand{\ppket}[3]{ | \, #1 \rangle \otimes  | \, #2 \rangle 
                        \otimes  | \, #3 \rangle } 
\newcommand{\pproj}[2]{\ket{#1}\bra{#1}\otimes\ket{#2}\bra{#2} }
\newcommand{\abs}[1]{ | \, #1 \,  |} 
\newcommand{\av}[1]{\langle\,#1\,\rangle}
\newcommand{\braket}[2]{\langle #1 \,|\, #2 \rangle}
\newcommand{\asred}{\preceq}
\newcommand{\asequ}{\approx}
\newcommand{\exred}{\le}
\newcommand{\exequ}{\equiv}
\newcommand{\tr}[2]{{\rm Tr}_{\rm \scriptscriptstyle #1}(#2)}
\newcommand{\PT}[1]{(#1)^{\rm T_B}}

\section{Introduction}
Quantum entanglement, first noted by  Einstein-Podolsky-Rosen (EPR) 
\cite{ein:pod:ros:35} and  
Schr$\ddot{\rm o}$dinger \cite{sch:35}, is one of the  
essential features of quantum 
mechanics. Its famous embodiment, the spin singlet
\begin{equation} 
\ket{\Psi\upp{AB}}= \frac{1}{\sqrt{2}}(\ket{\uparrow\upp{A}\downarrow\upp{B}} 
- \ket{\downarrow\upp{A}\uparrow\upp{B}})
\end{equation}
proposed  by Bohm \cite{boh:52}, has been  shown 
by  Bell \cite{bel:64}  to have  stronger correlations
than allowed by any local hidden variable theory.
The GHZ state \cite{gre:hor:zei:89,mer:90}
\begin{equation}
\ket{\Psi\upp{ABC}}= \frac{1}{\sqrt{2}}(\ket{\uparrow\upp{A}\uparrow\upp{B}
\uparrow\upp{C}} 
+ \ket{\downarrow\upp{A}\downarrow\upp{B}\downarrow\upp{C}}) 
\end{equation}
is a canonical three-particle example of quantum entanglement. Contradiction
between local hidden variable theories and quantum mechanics occurs
even for non-statistical predictions about the GHZ state 
\cite{gre:hor:zei:89,mer:90}, as opposed to the 
statistical ones for the EPR singlet.
These aspects of quantum mechanics have often been referred to as quantum 
nonlocality and form an  important aspect  of the study of the foundations
of quantum mechanics.

Recently it has been realized that quantum resources can be useful 
in information processing. Quantum entanglement plays a key role in 
many such applications like quantum teleportation 
\cite{ben:bra:cre:joz:per:woo:93}, superdense coding \cite{ben:wie:92}, 
quantum error correction \cite{got:97}, quantum key distribution 
\cite{ben:bra:84}, entanglement enhanced classical communication 
\cite{ben:fuc:smo:97}, quantum computational speedups \cite{deu:85}, 
quantum distributed computation \cite{gro:97} and 
entanglement enhanced communication complexity \cite{cle:bur:97}. 
In view of its central role   \cite{rev:phywor:98} in quantum 
information, it is imperative to have a qualitative 
as well as quantitative theory of it.
 
In the last few years much progress has been made in the study of 
bipartite pure- and mixed-state entanglement. In the rest of the 
introduction we mention some of these results  that are needed. 
In section \ref{sec:tripurent} we take a brief look at 
a recently proposed framework for quantifying tripartite (multipartite) 
pure-state entanglement \cite{pop:96}. Finally 
in section \ref{sec:res} we present the results: the first result
establishing the equivalence of the set of Schmidt decomposable states and 
the set of multiseparable
states provides support for the proposed pure state entanglement measure,
and the second result provides a necessary condition for the existence of
bound entanglement with positive partial transpose.

Now let us look at  some basic properties  of  entanglement.
\subsection{Entanglement Basics}

Entanglement is a property that only has meaning for  a multi-partite
system, i.e. one whose Hilbert space can be viewed as a product of two 
or more tensor factors corresponding to physical subsystems of the system.
In the EPR example, the two subsystems are the two spin 1/2 particles 
A and B that form the spin singlet. As a matter of convenience, 
we think about these  subsystems as belonging to different parties: 
Alice has subsystem A, Bob has subsystem B and so on. 
For  arbitrary systems, EPR singlets and GHZ states can be made meaningful
by labelling any two orthogonal states of  each party's subsystem as spin-up
 and spin-down respectively.

Operationally, {\em unentangled} or {\em separable} states are the ones
that can be made by the different parties with (at most) {\em classically 
coordinated local operations}, i.e. local operations by the parties, which are
coordinated by the exchange of classical information. Here, {\em local 
operations} include unitary transformations, additions of ancillas,
measurements and throwing away parts of the system, all performed  locally by 
one party
on his/her subsystem. Classical communication between parties 
is included because it allows for the creation of mixed states that are 
classically correlated but exhibit no quantum correlations. 

Thus, mathematically speaking,
a pure state $\ket{\Psi\upp{ABC...}}$ is separable iff it can be written 
as a tensor product of states belonging to different parties:
\begin{equation}
\ket{\Psi\upp{ABC...}} = \ppket{\phi\upp{A}}{\chi\upp{B}}{\psi\upp{C}} 
\otimes ... \enspace. 
\end{equation}
A mixed state $\rho\upp{ABC...}$ is separable iff it can be written as a 
sum of separable pure states \cite{wer:89}:
\begin{equation}
\rho\upp{ABC...} = 
\sum_i p_i \proj{\phi_i\upp{A}} \otimes \proj{\chi_i\upp{B}}
\otimes \proj{\psi_i\upp{C}} 
\otimes ... \enspace,
\end{equation} 
where the probabilities $p_i\ge0$ and $\sum_i p_i = 1$.
Finally, states that are not separable are said to be 
{\em entangled} or {\em inseparable}.  

Because classical communication between parties should not increase their 
quantum correlations, the expectation of any quantitative measure of 
entanglement should be 
non-increasing under classically coordinated local operations. In addition,
any such measure must be invariant under local unitary transformations, 
because they only correspond to another choice of local bases. Naturally, 
such a measure must be zero for any separable state. Also, it is natural to 
require such a measure to be additive for tensor products. To summarize,
the four requirements for a good measure of entanglement are
\cite{ved:ple:rip:kni:97}:
\begin{itemize}
\item Zero for separable states.
\item Invariant under local unitary transformations.
\item Non-increasing under classically coordinated local operations.
\item Additive for tensor products.
\end{itemize}

Since bipartite entanglement is the simplest case, let us review it next.
 
\subsection{Bipartite Entanglement}

For bipartite pure states it has been  shown 
\cite{ben:ber:pop:sch:96,lo:pop:97}
 that partial entropy is a good entanglement measure. It is equal,
 both to the state's 
{\em entanglement of formation} (the number of singlets asymptotically 
required to prepare the state, using only classically coordinated local 
operations), and the state's {\em distillable entanglement} (the number of 
singlets asymptotically preparable from the state using only classically 
coordinated local operations). Here partial entropy is the von Neumann 
entropy\footnote{The von Neumann entropy $S$ of a density matrix $\rho$ 
is defined 
to be the Shannon entropy $H$ of its eigenvalues, i.e. $S(\rho) =
 H(\{\lambda_i\}) = - \sum_i \lambda_i \log_2 (\lambda_i)$, where $\lambda_i$
are the eigenvalues of $\rho$. }
 of the reduced density matrix left after tracing out any one of 
the two parties. Mathematically we write this as,
\begin{equation}
E(\Psi{\upp{AB}}) = S(\rho\upp{A}) = S(\rho\upp{B}) \enspace, 
\end{equation}
where $\rho\upp{A}=\tr{B}{\proj{\Psi\upp{AB}}}$ and so on.
For mixed states partial entropy is no longer a good measure
since it can be nonzero for some separable states like the 
completely random state. 
A variety of apparently distinct entanglement measures for bipartite mixed 
states have been discussed, including entanglement of formation, 
distillable entanglement \cite{ben:ber:pop:sch:96,ben:div:smo:woo:96}, 
entanglement of assistance \cite{div:fuc:mab:smo:tha:uhl:98}, 
relative entropy entanglement \cite{ved:ple:rip:kni:97} 
and locally unitarily invariant parameters of the density
matrix \cite{lin:pop:97}. However, no measure has been proved to  satisfy all 
the properties required of a good measure.

Qualitatively, the set of inseparable states can be divided into two subsets:  
{\em distillable states\/} --- inseparable states that  have finite 
positive distillable entanglement --- and {\em bound entangled states\/} --- 
inseparable states that have zero distillable entanglement. 

The {\em partial transpose} of a density matrix can be used to formulate 
necessary conditions for separability and distillability, where
partial transpose $\PT{\rho\upp{AB}}$ of a density matrix  
$\rho\upp{AB}$ in the basis $\ket{i\upp{A}j\upp{B}}$ is given by
\begin{equation}
\bra{i\upp{A}j\upp{B}}\PT{\rho\upp{AB}}\ket{k\upp{A}l\upp{B}}= 
\bra{i\upp{A}l\upp{B}}\rho\upp{AB}\ket{k\upp{A}j\upp{B}} \enspace.
\end{equation}
The positivity\footnote{We say a matrix $A$ is positive iff 
all its eigenvalues are non-negative. This definition coincides with
that of non-negative matrices in mathematical literature.} of the partial 
transpose\footnote{Clearly, the partial transpose of a density matrix
is basis dependent, but its eigenvalues are not.} of a density matrix,
or equivalently the positivity of a density matrix under partial 
transposition (PPT) is a necessary condition\cite{per:96} for 
separability. Similarly, negativity\footnote{Here negativity of a hermitian 
matrix means that at least one of its eigenvalues is negative}
of the density matrix under partial 
transposition (NPT)  is a necessary condition \cite{hor:hor:hor:98} for 
distillability.

\begin{figure}[ptbh]
\epsfxsize=3.25in
\epsfbox{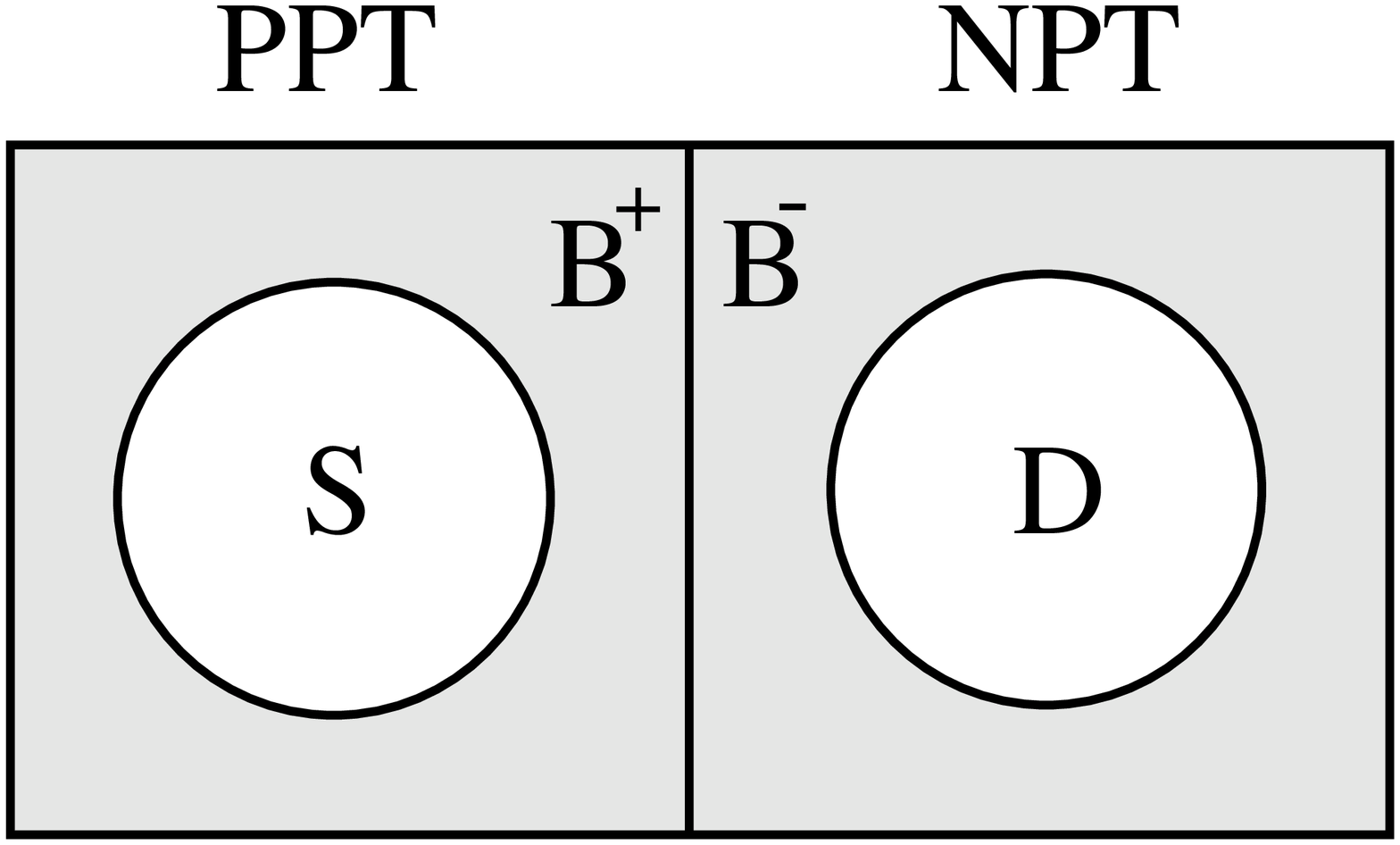}
\caption{{\bf Types of bipartite entanglement:}
 The left half of the figure represents the set of PPT states and 
the right half represents the set of NPT states. S denotes separable states 
and D denotes distillable states.  ${\rm B}^{+}$ (${\rm B}^{-}$) denotes 
bound entangled states with P(N)PT. In 
general, all these sets are known 
to be non-empty except for ${\rm B}^{-}$ for which no example is known yet.}
\label{fig:Typof2Ent}
\end{figure}

 Thus the set of mixed bipartite states can be divided into four classes 
as shown in 
Fig. \ref{fig:Typof2Ent}: the set of separable states (S), the set of 
distillable states (D), the set of PPT bound entangled
states (B$^+$) and the set of NPT bound entangled states (B$^-$).

Now we are in a position to turn to the concepts of reducibilities and
equivalences and their relation to entanglement measures. 

\section{Reducibilities, Equivalences and Entanglement Measures}
\label{sec:tripurent}

In this section we will review the concepts of reducibilities and 
equivalences with respect to classically coordinated local operations 
\cite{ben:97}, which are central to quantifying entanglement. Then we 
review a suggested way of quantifying tripartite and in general 
multipartite pure-state entanglement \cite{pop:96}.

In what follows we use a quantitative measure of similarity of two states.
One such measure is the fidelity \cite{uhl:76,joz:94}: the {\em fidelity} 
of a mixed state $\rho$ relative to a pure state $\ket{\psi}$ is given by
$F(\rho,\psi)= \bra{\psi}{\rho}\ket{\psi}$. It is the probability
with which $\rho$ will pass the test for being $\ket{\psi}$, conducted by
an observer who knows the state $\ket{\psi}$.

\subsection{Reducibilities and Equivalences} 
We say a pure state $\ket{\Phi}$ is {\em reducible} ($\exred$) to 
$\ket{\Psi}$ iff
\begin{equation}
\exists_{\cal L}\;\;\;{\cal L}(
\proj{\Psi})=\proj{\Phi}\enspace,
\end{equation}
where ${\cal L}$ is a  multi-locally implementable
trace preserving superoperator 
\cite{ben:div:fuc:mor:rai:sho:smo:woo:98,rai:98}
(a mathematical representation of classically coordinated local operations).
Intuitively this means that by doing 
classically coordinated local operations the parties can make
$\ket{\Phi}$ starting from $\ket{\Psi}$. Necessary and sufficient 
conditions for reducibility of bipartite pure states have been found 
in \cite{nie:98}.

Two states $\ket{\Phi}$ and $\ket{\Psi}$
are said to be {\em equivalent} ($\exequ$) iff $\ket{\Phi} 
\exred \ket{\Psi}$ and $\ket{\Psi} \exred \ket{\Phi}$. Intuitively this 
means that the two states are interconvertible by classically
coordinated local operations.
Here the principle of non-increase of entanglement implies that 
equivalent states must have the same entanglement \cite{ben:div:smo:woo:96}. 
Obviously, states related by local unitary transformations are 
equivalent and so are all separable states.

We say that 
$\ket{\Psi}^{\otimes x}$
and $\ket{\Phi}^{\otimes y}$, with $x$ and $y$ non-negative real numbers
\footnote{States with non-negative real exponents 
are defined by  $(\ket{\Psi}^{\otimes x})^{\otimes n} = 
\ket{\Psi}^{\otimes \lfloor{n\,x}\rfloor}$, where $\ket{\Psi}^{\otimes 0}=
 \ket{0\upp{A}0\upp{B}0\upp{C}...}$.}
, are {\em asymptotically equivalent} ($\approx$) 
iff $\ket{\Phi}^{\otimes y}$ is 
{\em asymptotically reducible ($\asred$)} to $\ket{\Psi}^{\otimes x}$ 
and vice versa, where 
\begin{eqnarray}
\ket{\Phi}^{\otimes y} \asred \ket{\Psi}^{\otimes x} \; \iff \; 
\forall_{\delta>0,\epsilon>0} \;\exists_{m,n,{\cal L}} \;\; 
\frac{\abs{m-n}}{m}<\delta & & \nonumber \\
{\rm and}\;\;F({\cal L}(\proj{\Psi}^{\otimes (m\,x)\ }),
\ket{\Phi}^{\otimes (n\,y)})  \ge 1-\epsilon \enspace & &.    
\end{eqnarray}
Here ${\cal L}$ is a multi-locally implementable trace preserving superoperator
that converts $m\,x$ copies of $\ket{\Psi}$ into a high fidelity approximation 
to $n\,y$ copies of $\ket{\Phi}$, where $m$ and $n$ are non-negative integers. 
These definitions extend the concepts 
of reducibility and equivalence to encompass the situation of 
asymptotic interconversion between states. Again the principle of non-increase
 of entanglement requires that asymptotically equivalent states must have the
 same entanglement.

As an example of the usefulness of these concepts let us 
re-express  the bipartite pure-state entanglement result 
\cite{ben:ber:pop:sch:96,lo:pop:97} mentioned in the introduction, 
in terms of 
asymptotic equivalence. In this new language, any bipartite pure state 
$\ket{\Psi\upp{AB}}$
is asymptotically equivalent to $E(\Psi\upp{AB})$ EPR singlets: 
\begin{equation}
\ket{\Psi\upp{AB}}\asequ \ket{{\rm EPR}\upp{AB}}^{\otimes E(\Psi\upp{AB})} 
\enspace.
\end{equation}
Thus if we take the EPR singlet to be the unit of entanglement (ebit), 
the partial entropy $E(\Psi\upp{AB})$
specifies the EPR singlets that can be obtained from and are required to 
prepare $\ket{\Psi\upp{AB}}$  by classically coordinated local operations.

In proving this result, the concepts of 
 entanglement concentration and 
dilution \cite{ben:ber:pop:sch:96} are central. The process of 
asymptotically reducing a given bipartite pure state to 
EPR singlet form is {\em entanglement dilution} and that of reducing 
EPR singlets to an arbitrary bipartite pure state is 
{\em entanglement concentration}. 
Then the above result means that entanglement concentration and dilution 
are reversible in the sense of 
asymptotic equivalence, i.e., in  the sense of approaching unit efficiency 
and fidelity in the limit of large number of copies $n$.  
The crucial
requirement for these methods to work is the existence of the Schmidt
(normal or polar) decomposition for bipartite pure states
\cite{hug:joz:woo:93}, in which any pure
state  say $\ket{\Psi\upp{AB}}$ can be written in the form
\begin{equation}
\ket{\Psi\upp{AB}}=\sum_i a_i \pket{i\upp{A}}
{i\upp{B}} \enspace,
\end{equation} 
where $\ket{i\upp{A}}$ and  $\ket{i\upp{B}}$ form 
orthonormal bases in Alice and Bob's 
Hilbert space respectively. Notice that, by change of phases  of local bases,
each of the {\em Schmidt coefficients} $a_i$ can be made real and 
non-negative. 

After this brief look at bipartite pure state entanglement,
let us look at the tripartite and multipartite case.

\subsection{Schmidt decomposability, multiseparability and  
pure-state entanglement}
\label{sec:triorth}

Let Alice, 
Bob, Charlie, ... , Nancy  be the  $n$ parties who have one subsystem each 
of an $n$-part system, generally in a joint state.

We say an $n$-party state $\ket{\Psi\upp{ABC...N}}$ is
($n$-){\em Schmidt decomposable} if it has an $n^{\rm th}$-order 
Schmidt decomposition, i.e. it can be written in the form:
\begin{equation}
\label{eq:nsch}
\ket{\Psi\upp{ABC...N}}=\sum_i a_i \ket{i\upp{A}}
\ket{i\upp{B}}\ket{i\upp{C}}...\ket{i\upp{N}} \enspace,
\end{equation} 
where $\ket{i\upp{A}}$, $\ket{i\upp{B}}$, $\ket{i\upp{C}}$, ..., 
$\ket{i\upp{N}}$ form 
orthonormal bases in Alice, Bob, Charlie, ..., and Nancy's  
Hilbert space respectively. Again, by change of phases  of local bases,
each of the {\em Schmidt coefficients} $a_i$ can be made real and 
non-negative. 

A useful property of Schmidt decomposable states is that the 
density matrices obtained by tracing out any 
party are separable. This property we call as {\em n-separability {\rm} or
multiseparability\/}\footnote{
In the terminology of \cite{bra:mor:98}, multiseparability is equivalent to
all possible partial semi-separability.}.
Further, these density matrices obtained by tracing one party are 
{\em eigenseparable\/}, i.e. they have separable
eigenvectors. This property we call as {\em n-eigenseparability {\rm or} 
multi-eigenseparability\/}.

Let us now look at tripartite states for concreteness.
It has been noted\cite{per:95} that arbitrary tripartite
pure states are not Schmidt decomposable. A different way of seeing this is by
using the facts that any bipartite
mixed state can be purified\footnote{A purification of a mixed 
state $\rho\upp{AB}$ is a  pure state $\ket{\Psi\upp{ABC}}$ such that 
$\rho\upp{AB}=\tr{C}{\proj{\Psi\upp{ABC}}}$.}
\cite{hug:joz:woo:93} to a tripartite pure state and  Schmidt decomposable
 states are tri-eigenseparable:
 Then if this were  true, all bipartite states would be 
eigenseparable, which is false.

Absence of a Schmidt decomposition for a general tripartite pure state 
means that, on
 the one hand the techniques developed for bipartite pure states cannot be 
generalized in a straightforward manner to tripartite pure states, but 
on the other hand it implies that  there are interesting new properties
 to be discovered about states that are not Schmidt decomposable. 

Now we are in a position to review the framework used \cite{pop:96} to 
quantify tripartite (multipartite) 
entanglement along the lines of the bipartite case. It is based on
 finding sets of states (analogous to the EPR singlet in the bipartite case)
 to which every pure tripartite (multipartite)  state is asymptotically
 equivalent.
Such sets are called reversibe entanglement generating sets. More precisely, 
set ${\cal G} = 
\{\ket{\psi_1},\ket{\psi_2},...,\ket{\psi_n}$\} is a {\em reversible
 entanglement generating set} (REGS) iff
for any state $\ket{\Psi}$ $\exists_{x_1,x_2,...,x_n\ge0}$,
such that 
\begin{equation}
\ket{\Psi} \asequ \ket{\psi_1}^{\otimes x_1} \otimes
 \ket{\psi_2}^{\otimes x_2} \otimes ... 
\otimes \ket{\psi_n}^{\otimes x_n} \enspace. 
\end{equation}
The tensor powers $x_1$, $x_2$,...,$x_n$ are known as 
the entanglement measure (or entanglement coefficients) induced by 
the REGS ${\cal G}$.

Of course one  would like to know the fewest states needed to make any 
general pure state.  This leads to the definition of a 
{\em minimal reversible entanglement  generating set} (MREGS) to be a REGS 
of least cardinality. 
The set ${\cal G}_2 =\{\ket{\rm EPR}\}$ is an 
example of a MREGS for bipartite entanglement which induces the 
entanglement measure given by the partial entropy in bits.

As mentioned earlier, bipartite entanglement concentration and
 dilution protocols depend crucially on the existence of a Schmidt
 decomposition. Not surprisingly, the bipartite protocols for entanglement 
concentration and dilution can be generalized \cite{pop:96} to work for
tripartite (multipartite) Schmidt decomposable states and used to prove 
that they are asymptotically  equivalent to GHZ (generalized  
GHZ\footnote{ An $n\/$-party generalization of the GHZ state which 
is called the {\em n-cat} state is defined to be,
$$\ket{n-{\rm cat}}=\frac{1}{\sqrt{2}}
(\ket{\uparrow\upp{A}\uparrow\upp{B}...\uparrow\upp{N}}+
\ket{\downarrow\upp{A}\downarrow\upp{B}...\downarrow\upp{N}}$$}) states,
 with the one-party partial entropy as the induced entanglement measure. 
That is, if $\ket{\Psi\upp{ABC}}$ is Schmidt decomposable,
\begin{equation}
\ket{\Psi\upp{ABC}} \asequ \ket{\rm GHZ}^{\otimes S(\rho\upp{A})} \enspace.
\end{equation}
We note here that for any multipartite Schmidt decomposable state,
one-party partial von Neumann entropies are equal to the Shannon entropy 
of the square of the Schmidt coefficients.

Now, we are in a position to motivate and present
 the main results of the paper.

\section{results}
\label{sec:res}

For simplicity let us  restrict ourselves to the case of tripartite systems.

The asymptotic equivalence of Schmidt decomposable states to GHZ states
gives a way of quantifying their entanglement. When we look at
states that do not lend themselves to the dilution and concentration scheme
for Schmidt decomposable states, we notice that it is their bipartite
entanglement left after tracing out a party, that somehow ``gets in the way'' 
of using these protocols. This fits in  with the fact we discussed earlier 
in section \ref{sec:triorth}
that the existence of non-Schmidt decomposable states is intimately
connected to the existence of bipartite entangled states. Thus it is
natural to expect that any triseparable state is Schmidt decomposable.

Here we prove this claim and in general prove that any multiseparable
state is Schmidt decomposable. Let us turn to that next.

\subsection{Equivalence of multiseparability and Schmidt decomposability}
\label{sec:tritri}

The result is trivial for one party. For the bipartite case it
says that all pure states have a Schmidt decomposition which 
as we mention earlier is known to be true. So, 
we prove the result first for the tripartite case and
then extend it to the multipartite case by induction.

Consider a triseparable pure state $\ket{\Psi\upp{ABC}}$. By definition,
$\rho\upp{AB}$, $\rho\upp{BC}$ and 
$\rho\upp{AC}$ are separable.

Now we show that  any triseparable state is Schmidt decomposable. 
Since PPT is a necessary (but in general not sufficient 
\cite{hor:hor:hor:96}) condition for separability, we prove a stronger 
result, namely:

{\em If a tripartite pure state $\ket{\Psi\upp{ABC}}$ is such that  
$\rho\upp{BC}$ is separable and $\rho\upp{AC}$ and  $\rho\upp{AB}$ are
PPT, then it is Schmidt decomposable. } 

\begin{figure}[ptbh]
\epsfxsize=3.25in
\epsfbox{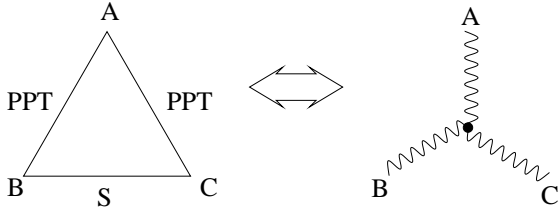}
\caption{{\bf Equivalence of Triseparability and Schmidt decomposability}:
Here, (A)lice, (B)ob, and (C)harlie represent the three parties. 
The sides AB, BC and AC of the triangle represent 
the density matrices $\rho\upp{AB}$, $\rho\upp{BC}$, and $\rho\upp{AC}$
respectively. 
 The wiggly lines represent ``essential''
tripartite entanglement embodied by Schmidt decomposable states.}
\label{fig:tseto}
\end{figure}

This result is illustrated in Fig. \ref{fig:tseto}.

To prove this we first write $\ket{\Psi\upp{ABC}}$ in its 
Schmidt decomposition \cite{hug:joz:woo:93} 
\begin{equation}
\label{eq:tstoe1}
\ket{\Psi\upp{ABC}}=\sum_{i=1}^{n}\sqrt{\lambda_i} 
                    \pket{\lambda_i\upp{A}}{\lambda_i\upp{BC}} \enspace ,
\end{equation} 
where $\ket{\lambda_i\upp{A}}$ are eigenvectors of $\rho\upp{A}$ and  
$\ket{\lambda_i\upp{BC}}$ are eigenvectors of $\rho\upp{BC}$ corresponding
to the non-zero (positive) eigenvalues $\lambda_i$.

Since $\rho\upp{BC}$ is separable it can be written as an ensemble 
of pure product states.  
Let ${\cal E}=\{p_i,\ket{\psi_i\upp{B}\phi_i\upp{C}}\;|\, i=1...m\}$   
be such an ensemble with the fewest members (here $m\ge n$).
Then probabilities $p_i> 0 \,, \forall_{i=1...m}$ and states 
$\ket{\psi_i\upp{B}\phi_i\upp{C}}$ are pairwise  linearly independent.  
Here $\ket{\psi_i\upp{B}\phi_i\upp{C}}$ is a short way of writing
$\pket{\psi_i\upp{B}}{\phi_i\upp{C}}$.

Now suppose Alice does the following local operations:
\begin{enumerate}
\item
She appends an ancilla and performs a local unitary transformation on 
$\ket{\Psi\upp{ABC}}$, resulting in
\begin{equation}
\label{eq:tstoe2}
\ket{\tilde{\Psi}\upp{ABC}} = \sum^m_{i=1} \sqrt{p_i} \ket{i\upp{A}
  \psi_i\upp{B}\phi_i\upp{C}} \enspace ,
\end{equation}
The Hughston-Jozsa-Wootters result \cite{hug:joz:woo:93} ensures that 
this is always possible.

\item
Now Alice chooses two distinct basis vectors $\ket{i\upp{A}}$ 
and $\ket{j\upp{A}}$,
 and does an incomplete von Neumann measurement projecting the above state
into the subspace spanned by these two vectors and its complement.
As a result, with probability $(p_i+p_j) > 0$ the joint state becomes 
\begin{equation}
\label{eq:tstoe4}
\nonumber
 \ket{\Psi\upp{ABC}_{ij}} = 
  q_i\ket{i\upp{A}\psi_i\upp{B}\phi_i\upp{C}} 
 + q_j\ket{j\upp{A}\psi_j\upp{B}\phi_j\upp{C}} 
\nonumber \enspace ,
\end{equation}
with $q_i=\sqrt{\frac{p_i}{p_i+p_j}}$ and $q_j=\sqrt{\frac{p_j}{p_i+p_j}}$.
\end{enumerate}

This can be rewritten as
\begin{eqnarray}
& &\ket{\Psi\upp{ABC}_{ij}}= q_i 
\ket{i\upp{A}i\upp{B}i\upp{C}} \nonumber \\
& & +q_j \ket{j\upp{A}} (\alpha \ket{i\upp{B}}+\
 \beta\ket{j\upp{B}}) (\gamma\ket{i\upp{C}}+\delta\ket{j\upp{C}})
\end{eqnarray}
where $\{\ket{i\upp{B}}, \ket{j\upp{B}}\}$  are orthonormal basis vectors 
for the $Span$ of $\{\ket{\psi_i\upp{B}},\ket{\psi_j\upp{B}}\}$
with $\ket{i\upp{B}}=\ket{\psi_i\upp{B}}$ 
and similarly  on Charlie's side. 
Also 
$\abs{\alpha}^2+\abs{\beta}^2=1$ and $\abs{\gamma}^2+\abs{\delta}^2=1$
for normalization. In this basis the
partial transpose of $\rho\upp{AB}_{ij}$ is
\begin{equation}
({\rho\upp{AB}_{ij}})^{\rm T\dnn{B}}= \left(
\begin{array}{cccc}
q_i^2 & 0 & q_iq_j\gamma^*\alpha^* & 0 \\
0 & 0 & q_iq_j\gamma^*\beta^* & 0\\
q_iq_j\gamma\alpha & q_iq_j\gamma\beta & 
q_j^2\abs{\alpha}^2 & q_j^2\alpha^*\beta  \\
0 & 0 &  q_j^2\alpha\beta^* & q_j^2\abs{\beta}^2 
\end{array}
\right) \nonumber \enspace .
\end{equation}
Since $\rho\upp{AB}$ is PPT so is $\rho\upp{AB}_{ij}$ \cite{hor:hor:hor:98};
 this requires
\cite{hor:joh:87}
$
\left| 
\begin{array}{cc}
0 & q_1q_2\gamma^*\beta^* \\
q_1q_2\gamma\beta & q_2^2\abs{\alpha}^2 
\end{array}
\right| \ge 0
$
, implying
\begin{eqnarray}
\gamma=0 & {\rm \; or \; } & \beta=0 {\rm \; \; i.e.}\nonumber \\
\ket{\phi_j\upp{C}} \perp \ket{\phi_i\upp{C}} & 
{\rm \; or \;} &\ket{\psi_j\upp{B}} = \ket{\psi_i\upp{B}}  \enspace .
\label{eq:tstoe5}
\end{eqnarray}

Repeating the above argument for $\rho\upp{AC}$ we get 
\begin{eqnarray}
\label{eq:tstoe6}
\ket{\psi_j\upp{B}} \perp \ket {\psi_i\upp{B}} &
{\rm \; or \;} & \ket{\phi_j\upp{C}} = \ket{\phi_i\upp{C}} \enspace. 
\end{eqnarray}

Since any pair of states in  ensemble ${\cal E}$ are linearly independent, 
the only consistent solution for the above equations is
\begin{eqnarray}  
\label{eq:tstoe7}
\ket{\psi_j\upp{B}}\; \perp\; \ket{\psi_i\upp{B}}& \;\;\; {\rm and}\;\;\; &
\ket{\phi_j\upp{C}}\; \perp\; \ket{\phi_i\upp{C}} \enspace .
\end{eqnarray}
Since Alice can  
choose any two distinct  $i,j=1...m$, Eq. (\ref{eq:tstoe7})  
implies that $\ket{\Psi\upp{ABC}}$ is 
Schmidt decomposable. This completes the proof. 
$\Box$

This result is  intuitively very satisfying, because it means that if 
there are no bipartite correlations among any two parties when the 
third party is traced out, then the tripartite state is Schmidt decomposable 
and hence asymptotically equivalent to GHZ states. This result supports 
the hypothesis that the GHZ and EPR states together form a MREGS, 
with the EPR singlets representing the bipartite entanglement between 
the parties and the GHZ state representing ``essential'' tripartite 
entanglement. 

The generalization of this result to the multipartite case follows 
by induction from the tripartite case. For convenience
we illustrate the induction step for the case of four parties: 
Alice, Bob, Charlie and David.

Let $\ket{\Psi\upp{ABCD}}$ be a 4-separable state of Alice, Bob, Charlie and 
David. By definition $\rho\upp{BCD}$, $\rho\upp{ACD}$, $\rho\upp{ABD}$, and 
$\rho\upp{ABC}$ are separable. Alice can by local operations as in the
paragraph before Eq. (\ref{eq:tstoe2}) make it into,
\begin{equation}
\label{eq:msmoe2}
\ket{\tilde{\Psi}\upp{ABCD}} = \sum^m_{i=1} \sqrt{p_i} \ket{i\upp{A}
  \psi_i\upp{B}\phi_i\upp{C}\chi_i\upp{D}} \enspace.
\end{equation}
Clubbing together Charlie and David into one party and applying the tripartite
result -- Eq. (\ref{eq:tstoe7}) -- implies that the $\ket{\psi_i\upp{B}}$ 
form an orthonormal set which we rename as $\ket{i\upp{B}}$. Thus,
\begin{equation}
\label{eq:msmoe3}
\ket{\tilde{\Psi}\upp{ABCD}} = \sum^m_{i=1} \sqrt{p_i} \ket{i\upp{A}
  i\upp{B}\phi_i\upp{C}\chi_i\upp{D}} \enspace.
\end{equation}
Now clubbing together Alice and Bob as a composite party 
$\widetilde{\rm Alice}$ and 
applying the tripartite result -- Eq. (\ref{eq:tstoe7}) --  we have that 
the $\ket{\phi_i\upp{C}}$ form an orthonormal set and so do the 
$\ket{\chi_i\upp{D}}$. This proves the result. $\Box$

After this we apply the two dimensional projection technique of this section
to prove a new necessary condition for bipartite bound entanglement with PPT.
 
\subsection{The No B$^+$-S theorem} 

It is well known that any two-party mixed state can be 
purified
 into a tripartite pure state. Then a connection between tripartite
pure-state entanglement and bipartite mixed-state entanglement seems likely. 
Already the fact that  triseparable states are Schmidt decomposable  tells us
 that a purification $\ket{\Psi\upp{ABC}}$ of a separable bipartite state 
$\rho\upp{AB}$ with inseparable  eigenvectors cannot be Schmidt decomposable
 and hence cannot be triseparable.
Thus, at least one of $\rho\upp{BC}$ and $\rho\upp{AC}$ is entangled. 

Here we prove another result of this kind: {any purification of a bipartite 
PPT bound-entangled (B$^+$) state is tri-inseparable}. 
More precisely,
 {\em if $\ket{\Psi\upp{ABC}}$ is a purification of a bipartite 
PPT bound-entangled state $\rho\upp{AB}$, then $\rho\upp{BC}$ 
and $\rho\upp{AC}$ are inseparable and hence $\ket{\Psi\upp{ABC}}$ is 
tri-inseparable. }

Before proving this result, note that any purification for $\rho\upp{AB}$
is related to any other, by addition of an ancilla and/or a local unitary 
transformation by Charlie \cite{hug:joz:woo:93}. Since inseparability 
is unaffected by such local operations \cite{ben:div:smo:woo:96},
if we prove
the above result for one purification it will hold for any other purification. 
The proof then follows as a trivial consequence of the following result.

{\em If a tripartite pure state $\ket{\Psi\upp{ABC}}$ is such that  
$\rho\upp{BC}$ is separable and $\rho\upp{AB}$ has positive partial
transpose, then $\rho\upp{AB}$ must be separable.   }

\begin{figure}[ptbh]
\epsfxsize=3.25in
\epsfbox{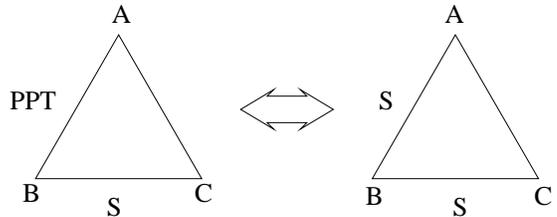}
\caption{{\bf No B$^+$-S Theorem}:
Here, (A)lice, (B)ob, and (C)harlie
represent the three parties. The sides AB, BC and AC of the triangles 
represent  the density matrices $\rho\upp{AB}$, $\rho\upp{BC}$, and
 $\rho\upp{AC}$ respectively.}
\label{fig:nobs1}
\end{figure}

This result is illustrated in Fig. \ref{fig:nobs1}.

The argument is similar to that employed in proving the equivalence of 
Schmidt decomposableity and triseparability. The difference 
is that here, only $\rho\upp{AB}$ is given to have positive partial 
transpose; but all the steps from Eq. (\ref{eq:tstoe1}) to 
Eq. (\ref{eq:tstoe5}) go through. 
To prove the result we show that Eq. (\ref{eq:tstoe5}) implies 
$\ket{\tilde{\Psi}\upp{ABC}}$ in Eq. (\ref{eq:tstoe2}) can be written as
\begin{equation}
\label{eq:nobs1}
\ket{\tilde{\Psi}\upp{ABC}} =\sum_{i=1}^{s} \ket{\mu_i\upp{B}} \otimes
\sum_{j=1}^{t(i)}\sqrt{p_{ij}}
\ket{\chi_{ij}\upp{A}} \otimes \ket{\nu_{ij}\upp{C}} \enspace ,
\end {equation}
with the kets
$\ket{\mu_i\upp{B}}$ pairwise linearly independent,
$\braket{\chi_{ij}\upp{A}}{\chi_{kl}\upp{A}}= \delta_{ik}\delta_{jl}$, and
$\braket{\nu_{ij}\upp{C}}{\nu_{kl}\upp{C}}=\delta_{ik} 
\braket{\nu_{ij}\upp{C}}{\nu_{il}\upp{C}}$. Here $\sum_{i=1}^{s} t(i)=m$
and $p_{ij} >0$ $\forall_{ij}$. 
Eq. (\ref{eq:nobs1}) implies,
\begin{eqnarray}
\tilde{\rho}\upp{AB}& = & {\rm Tr}\dnn{C} \proj{\tilde{\Psi}\upp{ABC}} 
\nonumber \\
& = & {\rm Tr}\dnn{C}\sum_{i=1}^s q_i \proj{\mu_i\upp{B}} 
     \otimes \proj{\chi_i\upp{AC}}
\end{eqnarray}
with $\ket{\chi_i\upp{AC}}=\sum_{j=1}^{t(i)} \sqrt{\frac{p_{ij}}{q_i}}
\ket{\chi_{ij}\upp{A}} \otimes \ket{\nu_{ij}\upp{C}}$ and 
$q_i=\sum_{j=1}^{t(i)} p_{ij}>0$; we have used the orthogonality of the 
$\ket{\nu_{ij}}$'s for different values of subscript $i$. 
Performing this trace 
it is easy to see that  $\tilde{\rho}\upp{AB}$ is separable. Recall that  
$\tilde{\rho}\upp{AB}$ is obtained  from $\rho\upp{AB}$ by appending an 
ancilla and/or a unitary rotation by Alice. Since inseparability is preserved 
under these local operations,  separability of $\tilde{\rho}\upp{AB}$ 
implies that $\rho\upp{AB}$ is also separable.

Now all that remains to be proved is that $\ket{\tilde{\Psi}\upp{ABC}}$ has
the form shown in Eq. (\ref{eq:nobs1}) above. For this we use induction on the 
number of terms in $\ket{\tilde{\Psi}\upp{ABC}}$. Obviously the form in 
Eq. (\ref{eq:nobs1}) holds in the case when  $\ket{\tilde{\Psi}\upp{ABC}}$
has just one term.
 Now assuming that 
this form holds for $s=r-1$ terms,
\begin{equation}
\ket{\tilde{\Psi}\upp{ABC}_r} = \sqrt{1-p_r} \ket{\tilde{\Psi}\upp{ABC}_{r-1}}
 + \sqrt{p_r} \ket{r\upp{A}\psi_r\upp{B}\phi_r\upp{C}} \enspace ,
\end{equation}
Where $\ket{\tilde{\Psi}\upp{ABC}\dnn{r-1}}$ has the form of 
Eq. (\ref{eq:nobs1}) with $\sum_{i=1}^{s}t(i)=r-1$. Here 
$1>p_r>0$.
But the condition in Eq. (\ref{eq:tstoe5}) with $j=r$ and $i=1...(r-1)$ 
implies that 
\begin{eqnarray}
{\rm either\;\;\;} \nonumber \\ 
& &\ket{\psi_r\upp{B}} = \ket{\mu_k\upp{B}} \; {\rm for}
 \; {\rm some} \; k 
\nonumber \\
& {\rm \;\;\;\;\;\;and\;\;\;\;\;} &  
\ket{\phi_r\upp{C}} \perp \ket{\nu_{ij}\upp{C}} \; \forall_{i,j;i\ne k} 
\nonumber\\
{\rm or\;\;\;} \nonumber \\
& & \ket{\phi_r\upp{C}} \perp \ket{\nu_{ij}\upp{C}} \; \forall_{i,j} 
\enspace .
\end{eqnarray}
Thus $\ket{\tilde{\Psi}_r\upp{ABC}}$ can be written in the form 
given by Eq. (\ref{eq:nobs1}); in the first case with 
 $s \to s$, $t(k)\to t(k)+1$ and  in the second case with, 
$s \to s+1$,  $t(s+1)=1$.
Thus the result is proved. $\Box$

\begin{figure}[ptbh]
\epsfxsize=3.25in
\epsfbox{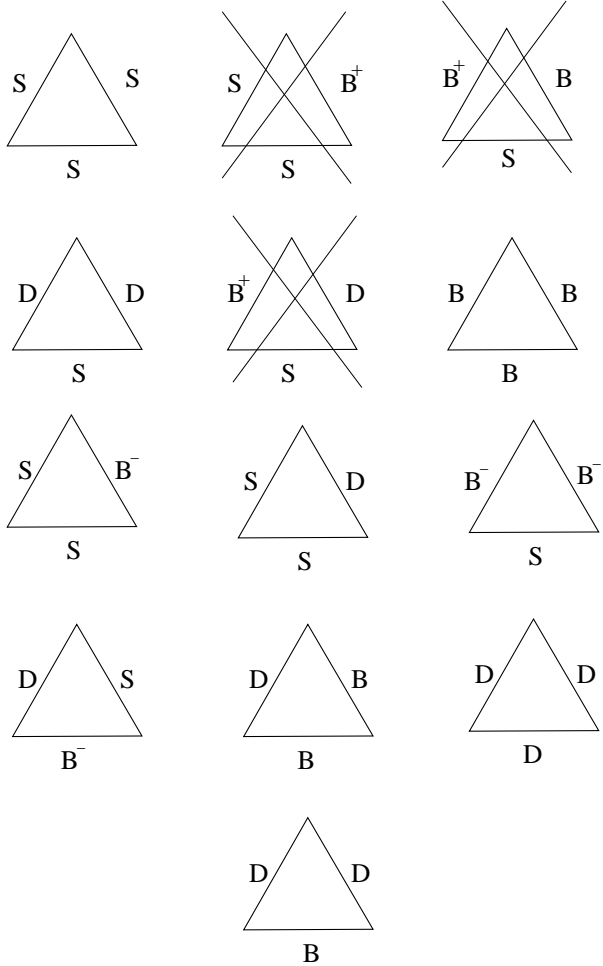}
\caption{Here vertices of the triangle represent the three parties, and 
each of the sides represents the corresponding bipartite density matrix
 obtained by tracing out the party corresponding to the remaining vertex. 
The letters near the sides label the kind of bipartite entanglement of the 
corresponding density matrices: (S)eparable, (D)istillable, B$^+$, B$^-$, 
and B, which stands for both B$^+$ and B$^-$.   }
\label{fig:NoBS}
\end{figure}

Given a tripartite pure state $\ket{\Psi\upp{ABC}}$,
there are many possibilities
for the kind of entanglement of the corresponding bipartite 
states $\rho\upp{AB}$, $\rho\upp{BC}$ and 
$\rho\upp{AC}$.
Figure \ref{fig:NoBS} shows these  possibilities and marks the 
ones ruled out by  this result. 

\section{Conclusions and Discussions}
We have proved the equivalence of multiseparable and Schmidt decomposable 
(multipartite pure) states. This result is relevant to
the problem of quantifying multipartite pure-state
entanglement, because it shows  that if 
there is no  ($n-1$)-partite entanglement after tracing any party,
then the $n$-partite state is 
Schmidt decomposable and hence asymptotically equivalent to the corresponding
generalized GHZ state, which
represents ``essential'' $n$-partite entanglement.
This result supports the hypothesis that the set of 2-,3-, ..., $n$-party 
generalized GHZ states form a minimal reversible entanglement generating set, 
with the $k$-party generalized GHZ states representing ``essential'' 
k-partite entanglement. Thus this work provides support for the 
entanglement measure proposed in \cite{pop:96}. 

We have also proved that any purification of a bipartite 
bound entangled state with positive partial transpose is tri-inseparable.
This provides a new necessary condition for bound entanglement with
positive partial transpose.

Further work needs to be done to prove that the generalized GHZ states 
form a minimal reversible entanglement generating set. 

An important question relating to the the No B$^+$S theorem is 
whether there exist states like ${\rm B}^+-{\rm B}^+-{\rm B}^+$
and ${\rm B}^+-{\rm B}^+-{\rm D}$. This 
is related to the question whether {\em tri-PPT} states are triseparable and
whether {\em bi-PPT} states are biseparable.

\section{Acknowledgments} 
 I would like to thank Charles H. Bennett, David P. DiVincenzo, Christopher 
A. Fuchs, John A. Smolin,
 Barbara Terhal, Armin Uhlmann, and William K. Wootters for
illuminating discussions.  I would
also like to thank David Awschalom for his invaluable support,
without which it would have been impossible to work in this exciting
field. 

Part of this work was completed during the 1998 Elsag-Bailey -- I.S.I.
Foundation research meeting on quantum computation. Travel support 
from The NSF Science and Technology Center for Quantized
Electronics Structures, Grant \#DMR 91-20007,
and logistic support from the IBM Research Division is greatly appreciated.
The support from the army research office under grant 
ARO DAAG55-98-1-0366 is greatly appreciated.

\end{document}